\documentclass[conference]{IEEEtran}
\usepackage{mathpazo}
\usepackage{times}
\usepackage{amsmath}
\usepackage{amsfonts}
\usepackage{latexsym}
\usepackage{amssymb}
\usepackage{cite}

\usepackage{upref}
\usepackage{amsthm}
\usepackage{graphicx}
\usepackage{psfrag}
\usepackage{subcaption}

\usepackage[usenames]{color}
\usepackage{overpic}

\usepackage{mathtools}
\usepackage{xparse}
\DeclarePairedDelimiterX{\set}[1]{\{}{\}}{\setargs{#1}}
\NewDocumentCommand{\setargs}{>{\SplitArgument{1}{;}}m}
{\setargsaux#1}
\NewDocumentCommand{\setargsaux}{mm}
{\IfNoValueTF{#2}{#1} {#1\,\delimsize|\,\mathopen{}#2}}

\DeclarePairedDelimiter\floor{\lfloor}{\rfloor}
\DeclarePairedDelimiter\parenv{\lparen}{\rparen}

\newcommand{\eps}{\varepsilon}

\theoremstyle{plain}

\newtheorem{theorem}{Theorem$\!$}

\newtheorem{lemma}{Lemma$\!$}

\newtheorem{corollary}{Corollary$\!$}

\newtheorem{definition}{Definition$\!$}

\newtheorem{remark}{Remark}

\newcounter{enumrom}
\renewcommand{\theenumrom}{(\roman{enumrom})}

\makeatletter
\renewcommand{\@endtheorem}{\endtrivlist}
\makeatother

\makeatletter
\renewcommand{\thefigure}{{\@arabic\c@figure}}
\renewcommand{\fnum@figure}{{\bf Figure\,\thefigure}}
\makeatother

\newcommand{\cN}{\mathcal{N}}

\newcommand{\cS}{\mathcal{S}}
\newcommand{\cT}{\mathcal{T}}

\newcommand{\be}[1]{\begin{equation}\label{#1}}
\newcommand{\ee}{\end{equation}}

\renewcommand{\leq}{\leqslant}

\renewcommand{\geq}{\geqslant}

\renewcommand{\Bbb}{\mathbb}

\newcommand{\Cref}[1]{Co\-ro\-lla\-ry\,\ref{#1}}

\renewcommand{\Bbb}{\mathbb}

\newcommand{\F}{{\Bbb F}}
\newcommand{\Fq}{{{\Bbb F}}_{\!q}}

\newcommand{\eqdef}{\triangleq}

\DeclareMathOperator{\rank}{rank}

\DeclareMathOperator{\gap}{gap_2}

\newcommand{\sbinom}[2]{\genfrac{[}{]}{0pt}{}{#1}{#2}}

\outer\def\proclaim #1. #2\par{\medbreak
 \noindent{\bf#1.\enspace}{\sl#2\par}%
 \ifdim\lastskip<\medskipamount \removelastskip\penalty55\medskip\fi}

\usepackage{pgf}
\usepackage{tikz}
\usetikzlibrary{matrix,chains,positioning,arrows,calc,decorations,plotmarks, patterns, fit,backgrounds}
\usepackage{pgfplots}
\usetikzlibrary{external,fillbetween,svg.path}
\usepackage{tkz-graph}
\pgfplotsset{compat=1.3}
\tikzstyle{help lines}=[black!20,dashed]

\pgfplotscreateplotcyclelist{mylist}{red,blue,black,yellow,brown}

\pgfdeclarepatternformonly{my north east lines}{\pgfqpoint{-1pt}{-1pt}}{\pgfqpoint{8pt}{8pt}}{\pgfqpoint{6pt}{6pt}}%
{
	\pgfsetlinewidth{0.4pt}
	\pgfpathmoveto{\pgfqpoint{0pt}{0pt}}
	\pgfpathlineto{\pgfqpoint{6.1pt}{6.1pt}}
	\pgfusepath{stroke}
}
\definecolor{light_gray}{rgb}{0.6,0.6,0.6}
\definecolor{awgray}{rgb}{0.7,0.7,0.7}
\definecolor{awgray_dark}{rgb} {0.4,0.4,0.4}

\tikzset{
	>=stealth',
	mycircle/.style={circle, draw=gray, very thick, text width=.1em, minimum height=1.5em, text centered},
	mycircle_small/.style={circle,draw=awgray_dark,fill = awgray_dark, inner sep=0,minimum size=.6em},
	mycircle_small_black/.style={circle,draw=black,fill = black, inner sep=0,minimum size=.6em},
	mybox/.style={rectangle,rounded corners,draw=black, thick,text width=1em,minimum height=4em,minimum width=4em,text centered},
	mybox_small/.style={rectangle,rounded corners,draw=black, thick,text width=1em,minimum height=2em,minimum width=2em,text centered},
	mybox_vec/.style={rectangle,rounded corners,draw=black, thick,text width=1em,minimum height=0.7em, minimum width=4em,text centered},
	mybox_vec_short/.style={rectangle,rounded corners,draw=black, thick,text width=1em,minimum height=0.7em, minimum width=2em,text centered},
	pil/.style={->, thick, shorten <=2pt, shorten >=2pt,},
}
\def\ve#1{{\mathchoice{\mbox{\boldmath$\displaystyle #1$}}%
              {\mbox{\boldmath$\textstyle #1$}}%
              {\mbox{\boldmath$\scriptstyle #1$}}%
              {\mbox{\boldmath$\scriptscriptstyle #1$}}}}

\newcommand{\A}{\ve{A}}
\newcommand{\B}{\ve{B}}
\newcommand{\x}{\ve{x}}
\newcommand{\y}{\ve{y}}

\newcommand{\quadbinom}[2]{\sbinom{#1}{#2}}

\usepackage{xcolor}
\definecolor{brightmaroon}{rgb}{0.76, 0.13, 0.28}
\definecolor{ao}{rgb}{0.0, 0.5, 0.0}
\definecolor{azure}{rgb}{0.0, 0.5, 1.0}
\definecolor{TUMBlue}{RGB}{0,101,189} 
\definecolor{TUMBlueDark}{RGB}{0,82,147} 
\definecolor{TUMBlueLight}{RGB}{152,198,234} 
\definecolor{TUMBlueMiddle}{RGB}{100,160,200} 

\definecolor{TUMElfenbein}{RGB}{218,215,203} 
\definecolor{TUMGreen}{RGB}{162,173,0} 
\definecolor{TUMOrange}{RGB}{227,114,34} 
\definecolor{TUMGray}{gray}{0.6} 

\definecolor{TUMGreenLight}{RGB}{0,124,48}
\definecolor{TUMRed}{RGB}{196,7,27}

\begin{document}


\IEEEoverridecommandlockouts
\title{\textbf{On the Gap between Scalar and Vector Solutions of Generalized Combination Networks}}

\author{
  \IEEEauthorblockN{ Hedongliang Liu\IEEEauthorrefmark{1}, Hengjia Wei\IEEEauthorrefmark{2}, Sven Puchinger\IEEEauthorrefmark{3}, Antonia Wachter-Zeh\IEEEauthorrefmark{4}, and Moshe Schwartz\IEEEauthorrefmark{5}}
  \IEEEauthorblockA{\IEEEauthorrefmark{1}Electrical and Computer Engineering, Technical University of Munich, \texttt{lia.liu@tum.de}}
  \IEEEauthorblockA{\IEEEauthorrefmark{2}Electrical and Computer Engineering, Ben-Gurion University of the Negev, \texttt{hjwei05@gmail.com}}
  \IEEEauthorblockA{\IEEEauthorrefmark{3}Applied Mathematics and Computer Science, Technical University of Denmark, \texttt{svepu@dtu.dk}}
  \IEEEauthorblockA{\IEEEauthorrefmark{4}Electrical and Computer Engineering, Technical University of Munich, \texttt{antonia.wachter-zeh@tum.de}}
  \IEEEauthorblockA{\IEEEauthorrefmark{5}Electrical and Computer Engineering, Ben-Gurion University of the Negev, \texttt{schwartz@ee.bgu.ac.il}\vspace{-2.0em}}
  \thanks{This research was supported by the  German  Research  Foundation (DFG) with a German Israeli Project Cooperation (DIP) under grant no.~PE2398/1-1, KR3517/9-1 and by the DFG Emmy Noether Program under grant No. WA3907/1-1.}
}

\maketitle
\begin{abstract}
We study scalar-linear and vector-linear solutions to the generalized combination network. We derive new upper and lower bounds on the maximum number of nodes in the middle layer, depending on the network parameters. These bounds improve and extend the parameter range of known bounds. Using these new bounds we present a general lower bound on the gap in the alphabet size between scalar-linear and vector-linear solutions.
\end{abstract}


\section{Introduction}
\label{sec:intro}
Network coding has been attracting increasing attention since the seminal papers~\cite{ACLY00,LiYeuCai03} as it increases the communication throughput compared to routing. An algebraic formulation for the network coding problem can be found in~\cite{KM03}. A detailed survey on multicast network coding can be found in~\cite{FraSol16}. 

Throughout this paper we consider only linear networks, namely, all the nodes in the network compute linear functions. In a line of recent works, there is a distinction between scalar network coding and vector network coding, depending on whether messages sent along network edges are scalars or vectors. Vector network coding was mentioned in~\cite{CDFZ06} as \emph{fractional network coding} and extended to vector network coding in~\cite{EbrFra11}. In~\cite{SunYanLonYinLi16}, a network was constructed whose minimal alphabet for a scalar linear solution is strictly larger than the minimal alphabet for a vector linear solution, albeit, this gap is just $1$. In~\cite{EW18}, a larger gap was found in certain carefully constructed networks, which was later  extended even to minimal networks in~\cite{CaiEtzSchWac19}.

The main object we study in this paper is the \emph{generalized combination network}. Originally, the (non-generalized) combination networks were first introduced in~\cite{RiiAhl06}. It was shown in~\cite{EW18,CaiEtzSchWac19} that vector linear network coding does not outperform scalar linear coding schemes in terms of the gap for \emph{minimal} combination networks. The \emph{generalized combination networks} were first introduced in~\cite{EW18} and a gap was shown to exist for certain network parameters.

The goal of this work is to investigate the gap between the minimum required alphabet size for scalar-linear and vector-linear solutions of generalized combination networks. Our main contributions are: we first develop new upper and lower bounds on the maximal number of nodes in the middle layer of such networks depending on the other parameters of the network. Our new upper bounds are better than a previous bound from~\cite{EZ19covering-multiple} in some parameter range and the lower bounds cover a wide range of network parameters. We then convert these bounds to bounds on the minimal alphabet size for a linear solution for many networks. Finally, we derive a lower bound on the gap for any fixed network structure. To the best of our knowledge, this is the first lower bound that applies to nearly all generalized combination networks.

The rest of this paper is organized as follows. In Section~\ref{sec:pre} we introduce the concept of generalized combination networks and provide the notation used throughout this paper. In Section~\ref{sec:ub_rs} we give two new upper bounds on the maximum number of middle-layer nodes, and in Section~\ref{sec:lb_rv} we give two new lower bounds on it. In Section~\ref{sec:bound_gap} we show the gap between the field sizes of scalar-linear and vector-linear solutions. In Section~\ref{sec:discussion} we conclude with a brief discussion of the results.

\section{Preliminaries}
\label{sec:pre}
\subsection{The Generalized Combination Network}
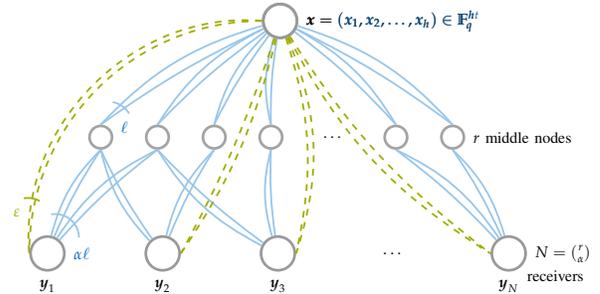
\begin{figure}[t]
  \centering
  \resizebox{.9\linewidth}{!}{\def\x{0.7}

\begin{tikzpicture}
  [font=\scriptsize,>=stealth',
  mycircle/.style={circle, draw=TUMGray, very thick, text width=.1em, minimum height=1.5em, text centered},
  mycircle_small/.style={circle,draw=TUMGray!90,very thick, inner sep=0,minimum size=1em,text centered},
  mylink/.style={color=TUMBlueLight, thick},
  mylink_dir/.style={color=TUMGreen, thick, dashed}]

  \coordinate (Source) at (0*\x,4*\x);
  {\node[mycircle,label=right:{$\ve{x}=({x}_1,{x}_2,\dots,{x}_h)\in\Fq^h$}] (Source) {};}
  {\node[mycircle,label=right:{$\ve{x}=\textcolor{TUMBlueDark}{(\ve{x}_1,\ve{x}_2,\dots,\ve{x}_h)\in\Fq^{ht}}$}] (Source) {};}

  \node[mycircle_small,below left = \x*60pt and \x*100pt of Source] (M0) {};
  \node[mycircle_small,right = \x*20pt of M0] (M1) {};
  \node[mycircle_small,right = \x*20pt of M1] (M2) {};
  \node[mycircle_small,right = \x*20pt of M2] (M3) {};
  \node[draw=none,right = \x*20pt of M3] (M4) {$\dots$};
  \node[mycircle_small,right = \x*20pt of M4] (M5) {};
  \node[mycircle_small,right = \x*20pt of M5] (M6) {};

  \node[mycircle,below left = \x*60pt and \x*20pt of M0,label=below:{$\ve{y}_1$}] (R0) {};
  \node[mycircle,right = \x*50pt of R0,label=below:{$\ve{y}_2$}] (R1) {};
  \node[mycircle,right = \x*50pt of R1,label=below:{$\ve{y}_3$}] (R2) {};
  \node[draw=none,right = \x*50pt of R2] (R3) {$\dots$};
  \node[mycircle,right = \x*50pt of R3,label=below:{$\ve{y}_{N}$}] (R4) {};

  \path[] (Source) edge[mylink,bend right=15] (M0.north)
  edge[mylink,bend right=15] (M1.north)
  edge[mylink,bend right=15] (M2.north)
  edge[mylink,bend right=15] (M3.north)
  edge[mylink,bend left=15] (M5.north)
  edge[mylink,bend left=15] (M6.north);

  \path[] (R0) edge[mylink,bend left=15] (M0.south)
  edge[mylink,bend left=15] (M1.south);
  \path[] (R1) edge[mylink,bend left=15] (M0.south)
  edge[mylink,bend left=15] (M2.south);
  \path[] (R2) edge[mylink,bend left=15] (M1.south)
  edge[mylink,bend left=15] (M3.south);
  \path[] (R4) edge[mylink,bend right=15] (M5.south)
  edge[mylink,bend right=15] (M6.south);

  \draw [mylink,solid] ($(M0)+(\x*18pt,\x*15pt)$) arc (30:80:\x*15pt); 
  \draw [mylink,-] ($(R0)+(\x*20pt,\x*10pt)$) arc (0:100:\x*15pt); 
  \node[draw=none,above right = \x*-8pt and \x*2pt of M0] (ell) {\textcolor{TUMBlue}{$\ell$}};
  \node[draw=none,right=0pt of M6] (rs) {$r$ middle nodes};
  \node[draw=none,right=0pt of R4] (N) {$N=\binom{r}{\alpha}$};
  \node[draw=none,below right=\x*-5pt and \x*-50pt of N] (recNodes) {receivers};

  \path[] (Source) edge[mylink,bend right=5] (M0.north)
  edge[mylink,bend right=5] (M1.north)
  edge[mylink,bend right=5] (M2.north)
  edge[mylink,bend right=5] (M3.north)
  edge[mylink,bend left=5] (M5.north)
  edge[mylink,bend left=5] (M6.north);

  \path[] (R0) edge[mylink,bend left=5] (M0.south)
  edge[mylink,bend left=5] (M1.south);
  \path[] (R1) edge[mylink,bend left=5] (M0.south)
  edge[mylink,bend left=5] (M2.south);
  \path[] (R2) edge[mylink,bend left=5] (M1.south)
  edge[mylink,bend left=5] (M3.south);
  \path[] (R4) edge[mylink,bend right=5] (M5.south)
  edge[mylink,bend right=5] (M6.south);

  \path[] (Source.west) edge[mylink_dir, bend right=45] (R0.west)
  edge[mylink_dir, bend right=50] (R0.west)
  (Source) edge[mylink_dir,bend left=5] (R1.east)
  edge[mylink_dir,bend left=10] (R1.east)
  (Source) edge[mylink_dir,bend left=15] (R2.east)
  edge[mylink_dir,bend left=20] (R2.east)
  (Source) edge[mylink_dir,bend right=15] (R4.west)
  edge[mylink_dir,bend right=10] (R4.west);

  \draw [mylink_dir,solid] ($(R0)+(-\x*15pt,\x*35pt)$) arc (100:60:\x*15pt); 
  \node[draw=none,above left = \x*10pt and 3pt of R0] (alphal) {\textcolor{TUMGreen}{$\varepsilon$}};
  \node[draw=none,right = 0pt of R0] (alphal) {\textcolor{TUMBlue}{$\alpha\ell$}};

\end{tikzpicture}}
  \caption{Illustration of $(\eps,\ell)-\mathcal{N}_{h,r,\alpha\ell+\eps}$ networks}\label{fig:Network}
  \vspace{-3ex}
\end{figure}
An $(\eps,\ell)-\cN_{h,r,\alpha\ell+\eps}$ generalized combination network is illustrated in Figure~\ref{fig:Network} (see also~\cite{EW18}). The network has $3$ layers. The first layer consists of a source with $h$ source messages. The source transmits $h$ messages to $r$ middle nodes via $\ell$ parallel links (solid lines) between itself and each middle node. Any $\alpha$ middle nodes in the second layer are connected to a unique receiver (again, by $\ell$ parallel links each). Each receiver is also connected to the source via $\eps$ direct links (dashed lines).

\subsection{Network Coding Solutions}

In the multicast setting, we have the following coding problem: for each node in the graph, find functions of its incoming messages to transmit on its outgoing links, such that each receiver can recover all the messages. Such an assignment of functions is called a \emph{solution} of the network. If these functions are linear, we obtain a \emph{linear solution}. In linear network coding, each linear function for a receiver consists of \emph{coding coefficients} for each incoming message. If the messages are scalars in $\Fq$ and the coding coefficients are vectors over $\Fq$, the solution is called a \emph{scalar linear solution}, denoted by $(q,1)$-{linear solution}. If the messages are vectors in $\Fq^t$, and the coding coefficients are matrices over $\F_q$, it is called a \emph{vector solution}, denoted by $(q,t)$-linear solution. 

It was shown in~\cite[Thm.~8]{EW18} that the $(\eps,\ell)-\cN_{h,r,\alpha\ell+\eps}$ network has a trivial solution if $h\leq\ell+\eps$ and it has no solution if $h>\alpha\ell+\eps$. In this paper we focus on the non-trivially solvable networks, so it is assumed $\ell+\eps< h \leq \alpha\ell+\eps$ throughout the paper.

\subsection{The Field Size Gap}
The field size of a linear solution is an important parameter that directly influences the complexity of the calculations at the network nodes. In order to investigate the advantage of vector solutions in terms of the field size, a metric to measure the improvement needs to be specified.
We follow the notations from~\cite{CaiEtzSchWac19} to distinguish between scalar and vector linear
solutions. Given a network $\cN$, let $$q_s(\cN):=\min\{q: \cN\textrm{ has a } (q,1)-\textrm{linear solution}\}.$$ 
The $(q_s(\cN),1)$ is said to be \emph{scalar-optimal}. 
Similarly, let $$q_v(\cN):=\min\{q^t: \cN\textrm{ has a } (q,t)-\textrm{linear solution}\}.$$ 
Note that $q_v(\cN)$ is defined by the size of the vector space, rather than the field size. 
For $q^t = q_v(\cN)$, a $(q,t)$-linear solution is called \emph{vector-optimal}.

By definition,
$q_s(\cN)\geq q_v(\cN).$
Small field sizes are preferable in practical algorithm designs for network coding~\cite{LSB2004,LS09,GSRMsep2019}.
We define the \emph{gap} as
\[ \gap (\cN)\eqdef \log_2 (q_s(\cN)) -\log_2(q_v(\cN)),\]
which intuitively measures the advantage of vector network coding by the amount of extra bits per link we have to pay for an optimal scalar-linear solution compared to an optimal vector-linear solution. We note that this differs from the definition of gap in~\cite{CaiEtzSchWac19}.

\subsection{Codes in the Grassmannian Space}

The Grassmannian $\mathcal{G}(n,k)$ is a set of all subspaces of $\mathbb{F}_q^n$ of dimension $k\leq n$. The cardinality of $\mathcal{G}(n,k)$ is the well-known $q$-binomial:
\begin{equation*}
  |\mathcal{G}(n,k)|=\quadbinom{n}{k}_q \triangleq \prod\limits_{i=0}^{k-1} \frac{q^n-q^i}{q^k-q^i}=\prod\limits_{i=0}^{k-1} \frac{q^{n-i}-1}{q^{k-i}-1},
\end{equation*}
where
\begin{equation}
    \label{eq:gauss}
q^{k(n-k)}\leq \quadbinom{n}{k}_q < \gamma \cdot q^{k(n-k)},
\end{equation}
with $\gamma\approx 3.48$~\cite[Lemma 4]{KK08}. 
\begin{definition}[Covering Grassmannian Codes~\cite{EZ19covering-multiple}]
  An $\alpha$-$(n,k,\delta)_q^c$ \emph{covering Grassmannian code} $\mathbb{C}$ is a subset of $\mathcal{G}(n,k)$ such that each subset with $\alpha$ codewords of $\mathbb{C}$ spans a subspace whose dimension is at least $\delta+k$ in $\mathbb{F}_q^n$.
\end{definition}
The following theorem from~\cite{EZ19covering-multiple} shows the connection between covering Grassmannian codes and linear network coding solutions.
\begin{theorem}[{\cite[Thm.~4]{EZ19covering-multiple}}]\label{thm:cover_scalar_sol}
  The $(\eps,\ell)-\mathcal{N}_{h,r,\alpha\ell+\eps}$ network is solvable with a $(q,t)$-linear solution if and only if there exists an $\alpha$-$(ht,\ell t,ht-\ell t-\eps t)_{q}^c$ code with $r$ codewords.
\end{theorem}

\section{Upper Bounds on the Middle Layer}
\label{sec:ub_rs}

In this section we fix the network parameters $\alpha,\ell,\eps, h$ and we upper bound the number of nodes in the middle layer,~$r$. 

\begin{lemma}\label{lm-full-t2}
Let $\alpha\geq 2$, $h,\ell,t\geq 1$, $\eps\geq 0$, $h-\eps \geq 2 \ell$, and let $\cT$ be a collection of subspaces  of $\F_q^{(h-\eps) t}$ such that
\begin{enumerate}
\item[(i)] each subspace has dimension at most $\ell t$; and
\item[(ii)] any subset of $\alpha$ subspaces spans $\F_q^{(h-\eps)t}$.
\end{enumerate}
Then  we have $\alpha \ell  \geq h-\eps$ and
$$|\cT| \leq \parenv*{\floor*{ \frac{h-\eps}{\ell} }-2} +\parenv*{\alpha- \floor*{ \frac{h-\eps}{\ell} }+1} \sbinom{\ell t+1}{1}.$$
\end{lemma}

\begin{IEEEproof}
Take arbitrarily $\lfloor \frac{h-\eps}{\ell} \rfloor-2$ subspaces  from $\cT$ and take arbitrarily a subspace  $W$ of dimension $(h-\eps)t-\ell t-1$ which contains all these $\lfloor \frac{h-\eps}{\ell} \rfloor-2$  subspaces. Then
for any subspace $T\in \cT$, there is a hyperplane of $\F_q^{(h-\eps)t}$ containing both $W$ and $T$.
Note that there are $\sbinom{\ell t+1}{\ell t}=\sbinom{\ell t +1}{1}$ hyperplanes containing $W$ and each of them contains at most $\alpha-1$ subspaces from $\cT$. Thus
\begin{align*}
 |\cT| &\leq  \parenv*{\floor*{ \frac{h-\eps}{\ell} }-2} \\ 
 &\quad\ + \sbinom{\ell t+1}{\ell t}\parenv*{\alpha-1 - \parenv*{\floor*{ \frac{h-\eps}{\ell} }-2}} \\
  &=  \parenv*{\floor*{ \frac{h-\eps}{\ell} }-2} +\parenv*{\alpha- \floor*{ \frac{h-\eps}{\ell} }+1}  \sbinom{\ell t+1}{1}.
\end{align*}
\end{IEEEproof}

\begin{theorem}\label{thm-upbound-v2}
Let $\alpha\geq 2$, $h,\ell,t\geq 1$, $\eps\geq 0$, $h-\eps\geq 2\ell$, and let $\cS$ be a collection of  subspaces of $\F_q^{ht}$ such that
\begin{enumerate}
\item[(i)] each subspace has dimension at most $\ell t$; and
\item[(ii)] any subset of $\alpha$ subspaces spans a subspace of dimension at least $(h-\eps)t$.
\end{enumerate}
Then  we have $\alpha \ell  \geq h-\eps$ and
\begin{align*}|\cS| &\leq \sbinom{(\eps+\ell)t}{\eps t} \parenv*{ \parenv*{\alpha- \floor*{ \frac{h-\eps}{\ell} }+1}  \frac{q^{\ell t+1}-1}{q-1}-1} \\
&\quad\ + \floor*{\frac{h-\eps}{\ell}}-1\\
&\overset{(*)}{<} \gamma\parenv*{\alpha- \floor*{ \frac{h-\eps}{\ell} }+1} q^{\ell t (\eps t+1)} {+ \floor*{ \frac{h-\eps}{\ell}}-1}.
\end{align*}
\end{theorem}

\begin{IEEEproof}
Take arbitrarily $\floor*{\frac{h-\eps}{\ell}}-1$ subspaces from $\cS$ and a subspace $W \subset \F_q^{ht}$ of dimension $(h-\eps)t-\ell t$ such that $W$ contains all these  $\floor*{\frac{h-\eps}{\ell}}-1$ subspaces.  Then for any subspace $S \in \cS$ there is a subspace of dimension $(h-\eps)t$ containing both $W$ and $S$.

Let $m\triangleq \sbinom{(\eps+\ell )t}{\eps t}$. Then there are $m$ subspaces of dimension $(h-\eps)t$ containing $W$, say $W_1, W_2, \ldots, W_m$. Note that every $\alpha$ subspaces in $W_i \cap \cS$ span the subspace $W_i$. According to Lemma~\ref{lm-full-t2}, we have
\begin{align*}
    |W_i \cap \cS| &\leq \parenv*{\floor*{ \frac{h-\eps}{\ell} }-2}
    +\parenv*{\alpha- \floor*{ \frac{h-\eps}{\ell} }+1}  \sbinom{\ell t +1}{1}.
\end{align*}
Hence,
\begin{align*}
|\cS|  &\leq \sum_{i=1}^{m}\parenv*{|W_i\cap \cS|- \parenv*{\floor*{ \frac{h-\eps}{\ell}}-1  }} + \floor*{ \frac{h-\eps}{\ell}}-1 \\
 &\leq  \sbinom{(\eps+\ell)t}{\eps t} \parenv*{ \parenv*{\alpha- \floor*{ \frac{h-\eps}{\ell} }+1}  \frac{q^{\ell t+1}-1}{q-1}  -1   } \\
 &\quad\ + \floor*{ \frac{h-\eps}{\ell}}-1.
 \end{align*}
The inequality $(*)$ is derived by \eqref{eq:gauss}.
\end{IEEEproof}
The following corollary rephrases Theorem~\ref{thm-upbound-v2} with network parameters.

\begin{corollary}\label{cor:imupperbound-N}
Let $\alpha\geq 2$, $h,\ell,t\geq 1$, $\eps\geq 0$, and $h-\eps \geq 2\ell$. If $(\eps,\ell)-\mathcal{N}_{h,r,\alpha\ell+\eps}$ has a $(q,t)$-linear solution then
\[
r 
< \gamma\theta q^{\ell t (\eps t+1)} +\alpha-\theta,
\]
where $\theta\triangleq \alpha- \floor*{ \frac{h-\eps}{\ell} }+1$.
\end{corollary}
\begin{IEEEproof}
If a $(q,1)$-linear solution exists, then each of the $r$ nodes in the middle layer gets a subspace of dimension $\ell t$ of the source messages space. Since all receivers are able to recover the entire source message space, every $\alpha$-subset of the middle nodes span a space of dimension at least $(h-\eps)t$. We then use Theorem~\ref{thm-upbound-v2}.
\end{IEEEproof}

Theorem~\ref{thm-upbound-v2} and Corollary~\ref{cor:imupperbound-N} are valid for all $\alpha\geq 2$. However, we derive a better upper bound for $\alpha = 2$, as shown in the following theorem. 

\begin{theorem}\label{thm:imupbound-2}
Let $h,\ell,t\geq 1$, $\eps\geq 0$, and let $\cS$ be a collection of  subspaces of $\F_q^{ht}$ such that
\begin{enumerate}
\item[(i)] each subspace has dimension at most $\ell t$; and
\item[(ii)] the sum of any two subspaces has dimension at least $(h-\eps)t$.
\end{enumerate}
Then we have
\[|\cS|  \leq \frac{\sbinom{ht}{2\ell t - (h-\eps)t+1}}{ \sbinom{\ell t}{2\ell t - (h-\eps)t+1}}
 \leq \gamma \cdot q^{(h-\ell)(2\ell+\eps-h)t^2+(h-\ell)t}. 
\]
\end{theorem}

\begin{IEEEproof}
We may assume that each subspace has dimension  $\ell t$. Since the sum of every two subspaces has dimension at least $(h-\eps)t$, then their intersection has dimension at most $2\ell t - (h-\eps)t$. It follows that any subspace of dimension $2\ell t - (h-\eps)t+1$ is contained in at most one subspace of $\cS$. Note that there are $\sbinom{ht}{2\ell t - (h-\eps)t+1}$ subspaces of dimension  $2\ell t - (h-\eps)t+1$  and each   subspace of dimension $\ell t$ contains $\sbinom{\ell t}{2\ell t - (h-\eps)t+1}$ such spaces.  We have that
$$|\cS| \leq \sbinom{ht}{2\ell t - (h-\eps)t+1}/ \sbinom{\ell t}{2\ell t - (h-\eps)t+1}.$$
\end{IEEEproof}

\section{Lower Bounds on the Middle Layer Nodes}
\label{sec:lb_rv}

We now turn to study a lower bound on the number of nodes in the middle layer, when we fix the network parameters $\alpha,\ell,\eps,h$. The main results are summarized in Theorem~\ref{thm:LLL_bound} and Corollary~\ref{cor:EK19_ub}. In the following, we first give the condition on the coding coefficients under which a linear solution exists.

Let $\x_1,\dots,\x_h\in\mathbb{F}^t_{q}$ denote the $h$ source messages and $\y_1,\dots,\y_N\in\mathbb{F}^{(\eps+\alpha\ell)t}_q$ the messages received by each receiver\footnote{The vector $\y_i$ is the concatenation of all the messages received by the $i$th receiver node.}. Since each middle-layer node receives $\ell$ incoming edges, and has $\ell$ outgoing edges directed at a given receiver, we may assume without loss of generality that this node just forwards its incoming messages. Let us denote the \emph{coding coefficients} used by the source node for the messages transmitted to the $r$ middle nodes by $\A_1,\dots,\A_r \in \Fq^{\ell t \times ht}$. Additionally, we denote the coding coefficients used by the source node for the messages transmitted directly to the receivers by $\B_1,\dots,\B_N\in\F_q^{\eps t\times ht}$.

Each receiver has to solve the following linear system of equations (LSE):
\begin{equation*}
\y_i
=\begin{pmatrix}
  \A_{i_1}\\ \vdots\\ \A_{i_\alpha}\\ \B_i
\end{pmatrix}_{(\eps+\alpha\ell)t\times ht}
\cdot
\begin{pmatrix}
\x_1\\
\vdots\\
\x_{h}
\end{pmatrix}_{ht\times 1},\ \forall i =1,\dots,N=\binom{r}{\alpha},
\end{equation*}
where $\{\A_{i_1},\dots, \A_{i_\alpha}\}\subset \{\A_1,\dots,\A_r \}$.

Any receiver can recover the $h$ source messages $\x_1,\dots,\x_h$ if and only if
\begin{equation}\label{eq:NC_sol}
\rank
\begin{pmatrix}
  \A_{i_1}\\ \vdots\\ \A_{i_\alpha}
\end{pmatrix}_{\alpha \ell t\times ht}
\geq (h-\eps)t,\ \forall i =1,\dots,N.
\end{equation}
Here the {solution} of the $(\eps,\ell)-\mathcal{N}_{h,r,\alpha\ell+\eps}$ network is a set of the coding coefficients $\{\A_1,\dots,\A_r \}$ s.t.~\eqref{eq:NC_sol} holds (where $\B_1,\dots,\B_N$ may be easily determined from the solution). 

\subsection{A Lower Bound by the Lov\'asz-Local Lemma}
\begin{lemma}[The Lov\'asz-Local-Lemma~{\cite[Ch.~5]{TheProbMethod}}\cite{LLLbeck1991}]\label{lem:LLL}
Let $\mathcal{E}_1, \mathcal{E}_2, \hdots,\mathcal{E}_k$ be a sequence of events. Each event occurs with probability at most $p$ and each event is independent of all the other events except for at most $d$ of them.
If $epd\leq 1$, then there is a non-zero probability that none of the events occurs. 
\end{lemma}
We choose the matrices $\A_1,\dots,\A_r \in \Fq^{\ell t \times ht}$ independently and uniformly at random.
For $1 \leq i_1 <  \dots<i_\alpha \leq r$, we define the event
\begin{align*}
\mathcal{E}_{i_1,\dots,i_\alpha} \triangleq \set*{ (\A_{i_1},\dots, \A_{i_\alpha}) ; \rank \begin{pmatrix}
\A_{i_1} \\ \vdots \\ \A_{i_\alpha}
\end{pmatrix} < (h-\eps)t }.
\end{align*}

Let $p=\Pr(\mathcal{E}_{i_1,\dots,i_\alpha})$ and denote by $d$ the number of other events $\mathcal{E}_{i'_1,\dots,i'_\alpha}$ that are dependent on $\mathcal{E}_{i_1,\dots,i_\alpha}$.

\begin{lemma}\label{lem:upper_bound_on_p}
Let $\alpha\geq 2$, $h,\ell,t\geq 1$, $\eps\geq 0$. Fixing $1 \leq i_1 < \dots < i_\alpha \leq r$, we have
\begin{align*}
\Pr(\mathcal{E}_{i_1,\dots,i_{\alpha}}) \leq 2\gamma \cdot q^{(h-\alpha\ell-\eps)\eps t^2+(h-\alpha\ell-2\eps)t-1}.
\end{align*}
\end{lemma}
\begin{IEEEproof}
The number of matrices $\A\in\mathbb{F}_q^{m\times n}$ of rank $s$ is
\begin{equation}
  M(m,n,s)\triangleq\prod\limits^{s-1}_{j=0}\frac{(q^m-q^j)(q^n-q^j)}{q^s-q^j}
  \leq \gamma\cdot q^{(m+n)s-s^2} \label{eq:number_matrices_upper_bound}.
\end{equation}

Then,
\begin{align}
  \Pr(\mathcal{E}_{i_1,\dots,i_{\alpha}}) &=\frac{\sum\limits^{(h-\eps)t-1}_{i=0}M(\alpha \ell t,ht,i)}{q^{\alpha \ell h t^2}}\nonumber\\
    & \leq \frac{\sum\limits^{(h-\eps)t-1}_{i=0}\gamma \cdot q^{(h+\alpha\ell)ti-i^2}}{q^{\alpha\ell h t^2}}\label{eq:NM_upper_bound} \\
    & \leq \gamma \cdot \frac{q}{q-1}\cdot q^{\max_i\{(h+\alpha\ell)ti-i^2\} - \alpha\ell h t^2} \label{eq:summation_upper_bound}\\
    & = \gamma \cdot \frac{q}{q-1}\cdot q^{(h+\alpha\ell)ti-i^2|_{i=(h-\eps)t-1} - \alpha\ell h t^2} \label{eq:quadratic_maximum}\\
    & \leq \gamma\cdot 2\cdot q^{(h-\alpha\ell-\eps)\eps t^2+(h-\alpha\ell-2\eps)t-1}
\end{align}
where (\ref{eq:NM_upper_bound}) holds due to~\eqref{eq:number_matrices_upper_bound}, (\ref{eq:summation_upper_bound}) follows from a geometric sum, and \eqref{eq:quadratic_maximum} follows by maximizing $(h+\alpha\ell)ti-i^2$.
\end{IEEEproof}

\begin{lemma}\label{lem:upper_bound_on_d}
Let $\alpha\geq 2$, $h,\ell,t\geq 1$, $\eps\geq 0$. Fixing $1 \leq i_1 <\dots < i_\alpha \leq r$, the event $\mathcal{E}_{i_1,\dots,i_\alpha}$ is statistically independent of all the other events $\mathcal{E}_{i_1',\dots,i_\alpha'}$ ($1 \leq i_1' < \dots < i_\alpha' \leq r$), except for at most $\alpha\binom{r-1}{\alpha-1}$
of them.
\end{lemma}

\begin{IEEEproof}
For $1 \leq i_1 < \dots < i_\alpha \leq r$ and $1 \leq i_1' <\dots < i_\alpha' \leq r$, the events $\mathcal{E}_{i_1,\dots,i_\alpha}$ and $\mathcal{E}_{i_1',\dots,i_\alpha'}$ are statistically independent if and only if $\{i_1,\dots,i_\alpha\} \cap \{i_1',\dots,i_\alpha'\} = \emptyset$. Thus, having chosen $1 \leq i_1 < \dots < i_\alpha \leq r$, there are at most
$\alpha\binom{r-1}{\alpha-1}$
ways of choosing an independent event.
\end{IEEEproof}
\begin{remark}
Lemma~\ref{lem:upper_bound_on_d} is a union-bound argument on the number of dependent events. The exact number is $\binom{r}{\alpha}-\binom{r-\alpha}{\alpha}$.
However the exact expression makes it harder to resolve everything for $r$ later so we use the bound here.
\end{remark}
\begin{theorem}\label{thm:LLL_bound}
  Let $\alpha\geq 2$, $\eps \geq 0$, $\ell,t \geq 1$, and $1\leq h \leq \alpha\ell+\eps$ be fixed integers. If
$r \leq \beta \cdot q^{\frac{f(t)}{\alpha-1}}$,
where $\beta \triangleq \parenv*{\frac{(\alpha-1)!}{2e\gamma\alpha}}^{\frac{1}{\alpha-1}}$ and $f(t)\triangleq(\alpha\ell+\eps-h)\eps t^2+(\alpha\ell+2\eps-h)t +{1}$, then $(\eps,\ell)-\mathcal{N}_{h,r,\alpha\ell+\eps}$ has a $(q,t)$-linear solution.
\end{theorem}

\begin{IEEEproof}
By the Lov\'asz Local Lemma, it suffices to show that $epd\leq 1$. Noting that $d\leq \alpha\binom{r-1}{\alpha-1}\leq \alpha\cdot \frac{(r-1)^{\alpha-1}}{(\alpha-1)!}$, we shall require
\begin{align*}
    e\cdot2\gamma q^{(h-\alpha\ell-\eps)\eps t^2+(h-\alpha\ell-2\eps)t-1}
    \cdot \alpha\frac{(r-1)^{\alpha-1}}{(\alpha-1)!}\leq 1.
\end{align*}
Namely,
$ r \leq {\beta}\cdot 
     q^{\frac{(\alpha\ell+\eps-h)\eps}{\alpha-1}t^2+\frac{\alpha\ell+2\eps-h}{\alpha-1}t+\frac{1}{\alpha-1}}+1.$
We omit the plus one for simplicity.
\end{IEEEproof}

\subsection{A Lower Bound by $\alpha$-Covering Grassmannian Codes}
Let ${\cal B}_q(n,k,\delta;\alpha)$ denote the maximum possible size of an $\alpha$-$(n,k,\delta)_q^c$ covering Grassmannian code.

Let $\A$ be a $k\times (n-k)$ matrix, and let $\ve{I}_k$ be a $k\times k$ identity matrix. The matrix $[\ve{I}_k\ \A]$ can be viewed as a generator matrix of a $k$-dimensional subspace of $\Fq^{n}$, and it is called the \emph{lifting} of $\A$.
When all the codewords of an MRD code are lifted to $k$-dimensional subspaces, the result is 
called \emph{lifted MRD code}, denoted by $\mathbb{C}^{MRD}$.
\begin{theorem}\label{thm:EK_1_ext}
Let $n,k,\delta$ and $\alpha$ be positive integers such that $1\leq \delta \leq k$, $\delta+k\leq n$ and $\alpha\geq 2$. Then 
$${\cal B}_q(n,k,\delta;\alpha) \geq (\alpha -1) q^{\max\{k,n-k\}(\min\{k,n-k\}-\delta+1)}.$$
\end{theorem}

\begin{IEEEproof} Let $m=n-k$ and $K=\max\{m,n-m\}(\min\{m,n-m\}-\delta+1)$.
Since $\delta \leq \min\{m,n-m\}$,  an $[m\times (n-m), K, \delta]_q$ MRD code $\mathbb{C}$ exists.
Let $\mathbb{C}^{MRD}$ be the lifted code of $\mathbb{C}$. Then $\mathbb{C}^{MRD}$ is a subspace code of $\Fq^n$, which contains $q^K$ $m$-dimensional subspaces as codewords and its minimum subspace distance is $2\delta$~\cite{SKK08}.
Hence, for any two distinct $C_1,C_2 \in \mathbb{C}^{MRD}$ we have
$\dim (C_1 \cap C_2)\leq m-\delta$.

Now, let $\mathbb{D}=\set*{C^{\perp} ; C \in \mathbb{C}^{MRD}}$. Take $\alpha -1$ copies of $\mathbb D$ and denote their multiset union as  $\mathbb{D}^{(\alpha)}$. We claim that $\mathbb{D}^{(\alpha)}$ is an $\alpha$-$(n,k,\delta)_q^c$ covering Grassmannian code. For each codeword of $\mathbb{D}^{(\alpha)}$, since it is the dual of a codeword in $\mathbb{C}^{MRD}$, it has dimension $n-m$, which is $k$. For arbitrarily $\alpha$ codewords $D_1,D_2,\ldots, D_\alpha$ of $\mathbb{D}^{(\alpha)}$, there exist $1\leq i < j\leq \alpha$ such that $D_i \not= D_j$. Let $C_i=D_i^{\perp}$ and $C_j=D_j^\perp$. Then $C_i$ and $C_j$ are two distinct codewords of $\mathbb{C}^{MRD}$. It follows that
\begin{align*}
    \dim\parenv*{\sum_{\ell=1}^\alpha D_\ell} & \geq \dim \parenv*{D_i +D_j }
    = n - \dim \parenv*{D_i^{\perp} \cap D_j^{\perp}}\\
    & = n - \dim \parenv*{C_i \cap C_j}
    \geq n-m+\delta = k+\delta.
\end{align*}
So far we have shown that $\mathbb{D}^{(\alpha)}$ is an $\alpha$-$(n,k,\delta)_q^c$ covering Grassmannian code. Then the conclusion follows since
\begin{align*}
|\mathbb{D}^{(\alpha)}|= & (\alpha-1)|\mathbb{D}|=(\alpha-1)|\mathbb{C}^{MRD}| \\
= & (\alpha-1) q^{\max\{k,n-k\}(\min\{k,n-k\}-\delta+1)}.
\end{align*}
\end{IEEEproof}

As a consequence, we have the following:

\begin{corollary}\label{cor:EK19_ub}
Let $\alpha\geq 2$, $h,\ell,t\geq 1$, $\eps\geq 0$, $h\leq 2\ell+\eps$.
If
$r\leq (\alpha-1)q^{g(t)}$,
where
 \begin{align*}
     g(t)&\triangleq\max\{\ell t,(h-\ell)t\}\\
     &\quad\ \cdot(\min\{\ell t, (h-\ell)t\}-(h-\ell-\eps)t+1) \\
     &=\begin{cases}\ell\eps t^2 +\ell t & h\leq 2\ell, \\ (h-\ell)(2\ell+\eps-h)t^2+(h-\ell)t & \text{otherwise.} \end{cases}
 \end{align*}
then $(\eps,\ell)-\mathcal{N}_{h,r,\alpha\ell+\eps}$ has a $(q,t)$-linear solution.
\end{corollary}

Note that Theorem~\ref{thm:LLL_bound} and Corollary~\ref{cor:EK19_ub} are both sufficient conditions on $r$ s.t.~a solution exists for $(\varepsilon,\ell)-\cN_{h,r,\alpha\ell+\varepsilon}$. Thus they can be regarded as lower bounds on maximum number of nodes in the middle layer.

\section{Bounds on the Field Size Gap}
\label{sec:bound_gap}

In previous sections, we discussed bounds on the maximum number of nodes in the middle layer. To discuss $\gap (\cN)$, we first need the following conditions on the smallest field size $q_s(\cN)$ or $q_v(\cN)$, for which a network $\cN$ is solvable. 
\begin{lemma}\label{lem:lb_q}
Let $\alpha\geq 2$, $r,h,\ell,t\geq 1$, $\eps\geq 0$. If $(\eps,\ell)-\mathcal{N}_{h,r,\alpha\ell+\eps}$ has a $(q,t)$-linear solution then
    \begin{align*}
        q^t \geq 
        \begin{cases}
        \parenv*{ \frac{r+\theta-\alpha}{\gamma\cdot\theta}}^{\frac{1}{\ell (\eps t+1)}} & h\geq 2\ell+\eps,\\
        \parenv*{ \frac{r}{\gamma(\alpha-1)}}^{\frac{1}{\ell(\eps t+1)}} & \text{otherwise,}
        \end{cases}
    \end{align*}
    where $\theta \triangleq \alpha-\floor*{\frac{h-\eps}{\ell}}+1$ and $\gamma\approx 3.48$.
\end{lemma}
\begin{IEEEproof}
    It follows from Corollary~\ref{cor:imupperbound-N} that for $h\geq 2\ell+\eps$,
    $q^t\geq \parenv*{ \frac{r+\theta-\alpha}{\gamma\cdot\theta}}^{\frac{1}{\ell (\eps t+1)}}$,
    so the first case follows. The second case may be derived from~\cite{EZ19covering-multiple} in a similar manner.
\end{IEEEproof}

\begin{lemma}\label{lem:ub_q}
Let $\alpha\geq 2$, $r,h,\ell,t\geq 1$, $\eps\geq 0$. There exists a $(q,t)$-linear solution to $(\eps,\ell)-\mathcal{N}_{h,r,\alpha\ell+\eps}$ when 
    \begin{align}\label{eq:qv_ub}
    q^t \geq
    \begin{cases} 
    \parenv*{\frac{r}{\beta}}^{\frac{(\alpha-1)t}{f(t)}} & h\geq2\ell+\eps \\
    \parenv*{\frac{r}{\alpha-1}}^{\frac{t}{g(t)}}   &\text{otherwise,}
    \end{cases}
    \end{align}
    where $\beta$ and $f(t)$ are defined as in Theorem~\ref{thm:LLL_bound}, and $g(t)$ is defined as in Corollary~\ref{cor:EK19_ub}.
\end{lemma}
\begin{IEEEproof}
    The proof is similar to that in Lemma~\ref{lem:lb_q} and the cases follow from Theorem~\ref{thm:LLL_bound} and Corollary~\ref{cor:EK19_ub} respectively.
\end{IEEEproof}
Lemma~\ref{lem:lb_q} and Lemma~\ref{lem:ub_q} can be seen as the necessary and the sufficient conditions respectively on the pair $(q,t)$ s.t.~a $(q,t)$-linear solution exists. 

In the following, we use the lemmas above to derive a lower bound on the $\gap(\cN)$ for a given network $\cN$. The bound is determined only by the network parameters.
\begin{theorem}
\label{thm:gap_lb}
  Let $\alpha\geq 2$, $r,h,\ell\geq 1$, $\eps\geq 0$. Then for the $(\eps,\ell)-\mathcal{N}_{h,r,\alpha\ell+\eps}$ network,
  \begin{align*}
      \gap(\cN)\geq 
      \begin{cases}
      \frac{1}{\ell(\eps+1)}\log_2\parenv*{\frac{r+\theta-\alpha}{\gamma\theta}}-t_{\Delta} & h\geq 2\ell+\eps\\
      \frac{1}{\ell(\eps+1)}\log_2\parenv*{\frac{r}{\gamma(\alpha-1)}}-t_{\star} & \text{otherwise},
      \end{cases}
  \end{align*}
  where $t_{\Delta}$ is the smallest positive integer s.t.~$2^{\frac{f(t_{\Delta})}{\alpha-1}}\geq \frac{r}{\beta}$ and $t_{\star}$ is the smallest positive integer s.t.~$2^{g(t_{\star})}\geq \frac{r}{\alpha-1}$.
  Here, $\beta$ and $f(t)$ are defined as in Theorem~\ref{thm:LLL_bound}, and $g(t)$ is defined as in Corollary~\ref{cor:EK19_ub}.
\end{theorem}
\begin{IEEEproof}
 Let us first consider the first case $h\geq 2\ell+\eps$. According to Lemma~\ref{lem:lb_q}, we have the lower bound on the smallest field size of a scalar solution,
 $q_s(\cN)\geq \parenv*{ \frac{r+\theta-\alpha}{\gamma\cdot\theta}}^{\frac{1}{\ell (\eps +1)}}$,
For vector solutions, according to Lemma~\ref{lem:ub_q}, we want to find $(q,t)$ s.t.~$q^{\frac{f(t)}{\alpha-1}}\geq \frac{r}{\beta}.$
Since $t_{\Delta}$ is the smallest positive integer $t$ s.t.~$2^{\frac{f(t)}{\alpha-1}}\geq \frac{r}{\beta}$, it is guaranteed that a $(2,t_{\Delta})$-linear solution exists. 
Therefore, $q_v(\cN)$ (the smallest value of $q^t$) should be at most 
$q_v(\cN)\leq 2^{t_{\Delta}}$.
The lower bound then follows directly from the definition of $\gap(\cN)$. The other case can be proved in the same manner.
\end{IEEEproof}

By carefully bounding $t_{\star}$ and $t_{\Delta}$, the following is obtained:

\begin{corollary}
\label{cor:gap}
Let $\alpha\geq 2$, $r,h,\ell,\eps\geq 1$. Then for the $(\eps,\ell)-\mathcal{N}_{h,r,\alpha\ell+\eps}$ network,
\[
\gap(\cN) \geq
\begin{cases}
\frac{\log_2\left(\frac{r}{\alpha-1}\right)-2}{\ell(\eps+1)}-\sqrt{\frac{\log_2(\frac{r}{\alpha-1})}{\ell\eps}} & h\leq 2\ell+\eps,\\
\frac{\log_2\left(\frac{r+\theta-\alpha}{\gamma \theta}\right)}{\ell(\eps+1)} - \sqrt{\frac{(\alpha-1)\log_2(\frac{r}{\beta})}{(\alpha\ell+\eps-h)\eps}} & \text{otherwise.}
\end{cases}
\]
In particular, if all parameters are constants except for $r\to\infty$, then $\gap(\cN)=\Omega(\log r)$.
\end{corollary}

\section{Discussion}\label{sec:discussion}

In this work, we studied necessary and sufficient conditions for the existence of $(q,t)$-linear solutions to the generalized combination network. The derived conditions led us to find a lower bound on the gap for almost all network parameters. Unlike previous works, e.g., \cite{EW18,CaiEtzSchWac19}, which were focused on engineering specific networks with a high gap, we start with almost any given network, and provide an expression for its gap. It is of particular interest to note the implications of~Corollary~\ref{cor:gap}. Fixing the number of messages, and parameters relating to the connectivity level of the network, we only vary the number of middle layer nodes, $r$, or equivalently, the number of receivers $N\triangleq\binom{r}{\alpha}$. Corollary~\ref{cor:gap} then shows that the gap is $\Omega(\log r)=\Omega(\log N)$, namely, that scalar-linear solutions over-pay an order of $\log(r)$ extra bits per link to solve the network, in comparison with vector-linear solutions.
Our bounds, however, are weak in the case of no direct links, i.e., $\eps=0$, and improving them is left for future research.

In the full version of this paper we further study the bounds presented here, and compare them with the other known bounds of~\cite{EZ19covering-multiple}.

\bibliographystyle{IEEEtranS}
\bibliography{allbib}

\end{document}